\documentclass[usenatbib]{mn2e}

\usepackage{color}
\usepackage{graphicx}
\usepackage{epsf}
\usepackage{epsfig,graphicx,psfrag,amsfonts,amssymb}
\usepackage{amsmath}
\usepackage{dcolumn}
\begin{document} 
\title[]
{Thawing Versus. Tracker Behaviour: Observational Evidence}
\author[]
{Shruti Thakur$^{1}$, Akhilesh Nautiyal$^{2}$,  Anjan A Sen$^3$, T R Seshadri $^1$\\
$^1$ Department of Physics and Astrophysics, University of Delhi, Delhi-110007, India\\
$^2$ Harish-Chandra Research Institute, Allahabad, India\\
$^3$  Center for Theoretical Physics, Jamia Millia Islamia, New Delhi-110025, India}

\date{\today}
\maketitle
\pagerange{\pageref{firstpage}--\pageref{lastpage}} \pubyear{2010}

\begin{abstract}
Currently there are a variety of scalar field models that attempt to explain the late time acceleration of the Universe.  This includes the standard canonical and non-canonical scalar field models together with recently proposed Galileon scalar field models. One can divide all these scalar field models into two broad categories, namely the thawing and the tracker classes.
In this work we investigate the evidence for these models with the presently available  observational data using the Bayesian approach.
We use the Generalized Chaplygin Gas (GCG) parametrization for dark
energy equation of state (EoS) as it exhibits both the thawer
and tracker like behaviours for different values of the
parameters. Subsequently we use the SnIa ( SN) measurements, the recent measurements of the Hubble parameter at different redshifts (H(z)), measurements of the lookback time at different redshifts (Lookback), measurements of the linear growth  factor in large scale structure (GR) from various redshift surveys and finally the measurement of the anisotropies in the cosmic microwave background radiation by WMAP-7 observations. The analysis of data from SN+H(Z)+Lookback does not favour either the tracker or thawer classes. Inclusion of data from GR+WMAP-7 favours the thawer class of models if one assumes the dark energy to be a smooth component. But if we consider the dark energy perturbation, both tracker and thawer type of models are equally favoured by the data.
\end{abstract}


\section{Introduction}

It is now firmly established that the present Universe is going
through a phase of accelerated expansion. This has been confirmed by
different observational results including Type-Ia supernovae (\ Riess
et al.~(1998), \ Perlmutter et al.~(1999), \ Tonry et al.~(2003), \
Knop et al.~(2003), \ Riess et al.(2004), Amanullah et al.~(2010), \
Suzuki et al.~(2011)), cosmic microwave background radiation (\
Komatsu et al.~(2011)) as well as the latest surveys of the large
scale structue (\ Eisenstein et al.~(2005)).  The challenge is now to
explain the possible cause for this late time acceleration and there
has been a wide range of proposals for this. One of the possible sources for this accelerated expansion could be an exotic type of matter with an equation of state parameter $w < -1/3$. Such sources are collectively called Dark Energy. [see (\ Bean et al.~(2005), \ Copeland et al.~(2006), \ Li et al.~(2011), \ Padmanabhan(2003), \ Peebles \& Ratra ~(2003), \ Sahni \& Starobinsky~(2000)) for  review on dark energy models]. 

The simplest example for this dark energy is a cosmological constant
(CC) $\Lambda$, but the need for an extreme fine tuning and the cosmic
coincidence problem make $\Lambda$ an unattractive choice from a
theoretical perspective. The scalar field models are a natural
alternative. Different scalar field models including quintessence (\
Ratra \& Peebles~(1988),\ Caldwell et al.~(1998),\ Liddle \&
Scherrer~(1999),\ Steinhardt et al.~(1999)), tachyon (\ Abramo \&
Finelli~(2003),\ Bagla et al.~(2003),\ Aguirregabiria \&
Lazkoz~(2004),\ Copeland et al.~(2005)), phantom  (\ Caldwell~(2002))
and k-essence  (\ Armendariz et al.~(2001),\ Scherrer~(2004), \
Sen~(2006), \ Vikman~(2005)) as well as Galileons (\ Nicolis et
al.~(2009), \ Ali et al.~(2010), \ Gannouji \& Sami~(2010), \ Appleby
\& Linder~(2011), \ Linder~(2012), \ De. Felice \& Tsujikawa~(2010))
have been thoroughly investigated in recent time to explain the late
time acceleration of the Universe. In addition to circumvent the
problem like cosmic coincidence, the advantage of using these fields
is that they also mimic CC at late times. Moreover, as the equation of
state (EoS) for these scalar fields in general vary with time, it is
expected that these scalar field models can be distinguished from
$\Lambda$CDM with observational data from higher redshifts which are expected to become available in near future.

Scalar field models in general,  can be divided into two broad classes: the fast roll and slow roll models, also termed as tracking  and thawing models, respectively  (\ Caldwell \& Linder~(2005),\ Linder~(2003)). Tracker models with steep potentials usually mimic the background matter/radiation in the early time and remains subdominant. Only at late times, the field leaves the tracker regime and starts behaving like a component with large negative pressure and thereby behaves like  dark energy. On the other hand, thawing models are similar to inflaton, where the scalar field is initially frozen due to the large Hubble friction in the flat part of the potential and hence behaves close to CC.  At late times, the scalar field slowly thaws from this frozen state and departs from this CC behaviour. In both these cases, a certain degree of fine tuning (whether in the form of the potentials or in the initial conditions) is always necessary to get the required late time evolution.

With the high level of precision that exists in current observational
techniques, the question naturally arises as to whether the available
observational data indicate any clear preference of one of these
models over the other. In the present exercise, we attempt to answer
this question using the Bayesian approach. To do so, it is important
to look for a parametrization for the dark energy behaviour which
interpolates between the tracker and the thawer behaviour. It has
  been however pointed out that general barotropic models described by
  $p = p(\rho)$ can not fully mimic the scalar field models and in particular the tracker type of models (\ Scherrer~(2006), \ Linder \& Scherrer~(2009)). However, the attempt in this paper is to have a single model that broadly mimics the two overall behaviours, namely: 1)  the equation of state is frozen initially at $w=-1$ and then at late times it slowly thaws out from this frozen state which is similar to thawer type of scalar fields and 2) the equation of state is that of matter initially and later on starts behaving like cosmological constant which is similar to tracker type of scalar fields.
Fortunately a widely studied form,  known as "{\it Generalized Chaplygin Gas}" (GCG) (\ Kamenshchik et.al.~(2001), \ Bento et al.~(2002), \ Sen \& Scherrer~(2005)), can provide such a parametrization scheme [ See also (\ Aviles \& Cervantes-Cota (2011))]. This single parametrization for the EoS of the dark energy can not only mimic both of these two behaviours, it can also act as dark energy with constant EoS (including $\omega = -1$ which is the same as CC). Using this parametrization we show that considering only the homogenous background cosmology and applying it to analyze observational data does not discriminate between thawing and tracker type of behaviours. However, if we include the data from the growth of matter perturbations  as well as those from CMBR anisotropies, thawing type models are preferred if we do not consider the perturbation in the GCG fluid. But including perturbations in the GCG fluid results no preference of thawer over tracker type of behaviour. 

\begin{figure}
\includegraphics[width=8cm]{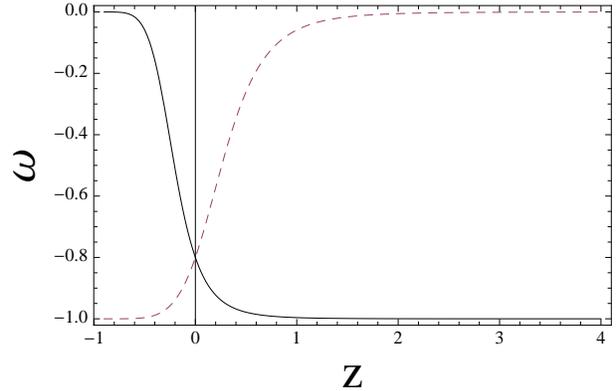}
\caption{Behaviour of the equation of state for GCG $\omega_ {_{ (gcg)}}$ as a function of redshift. The Solid line is for $(1+\beta) <0$ whereas the dashed line is for $(1+\beta) > 0$. $A_{s} = 0.85$ is assumed.}
\end{figure}

\section {GCG}
We start by discussing the equation of state for GCG given by,
\begin{equation}
p_ {_{ (gcg)}}=\frac{-A}{\rho_{_ { (gcg)}}^{\beta}},
\end{equation}
where $A$ and $\beta$ are two arbitrary constants (\ Bento et al.~(2002)). Assuming Friedmann-Robertson-Walker metric, this equation of state implies that the density evolves with redshift as
\begin{equation}
\rho_ {_{ (gcg)}}=\rho_{_{ (gcg)0}}\left[A_{s}+(1-A_{s})(1+z)^{3(1+\beta)}\right]^{\frac{1}{1+\beta}},
\end{equation}
where 
\begin{equation}
A_{s} = A [\rho_{_{ (gcg)0}}]^{-{(1+\beta)}} ,
\end{equation}
with $\rho_{_{ (gcg)0}}$ being the present value of $\rho_ {_{_{ (gcg)}}}$. The choice of the constants, $A_{s}$ and $\beta$ uniquely determines the dark energy behaviour in the GCG model.

One can easily show that $A_{s} = -w_ {_{ (gcg)}}(z=0)$. As our primary goal is to compare the thawing and tracking behaviours which primarily arise in the scalar field models, we confine ourselves to non-phantom region i.e $A_{s}\leq 1$.
For $A_{s} = 1$, GCG exactly behaves as  CC at all times and in this case the parameter $\beta$ is redundant. For $ A_{s} = 0$, GCG behaves exactly as a non-relativistic pressureless matter fluid at all times.

For $0<A_{s}<1$, the parameter $\beta$ determines the how the equation of state changes with time. For $\beta=-1$, the GCG behaves as a fluid with constant equation state $p_ {_{ (gcg)}}=- A \rho_ {_{ (gcg)}}$. For $(1+\beta) > 0$, at early times GCG behaves as matter while at late times it behaves as a fluid with negative equation of state and asymptotically tends to a Cosmological Constant like behaviour in the future.
This is exactly what we need for a tracking behaviour. On the other hand,  for $(1+\beta) < 0$, GCG behaves as a cosmological constant to start with and then as the Universe evolves, it starts deviating from this CC behaviour (\ Sen \& Scherrer~(2005)). This is similar to the thawing behaviour. Further, in this case, GCG always approaches to matter behaviour assymptotically in the future ($z= -1$). Hence, in this model, accelerated expansion is a transient feature. All these different behaviours are shown in Figure 1.

To summarise, GCG behaves as a tracking model for $(1+\beta) > 0$ whereas it behaves as thawing model for $(1+\beta) < 0$.  For $\beta = -1$,  GCG behaves as a fluid with a constant equation of state
\begin{equation}
w_{_{(gcg)}}=\frac{p_{_{(gcg)}}}{{\rho}_{_{(gcg)}}}=-A.
\end{equation}
It demarcates the thawing from the tracking regime.

\begin{figure*}
\begin{center}
\begin{tabular}{|c|c|}
\hline
 & \\
{\includegraphics[width=2.6in,height=2in,angle=0]{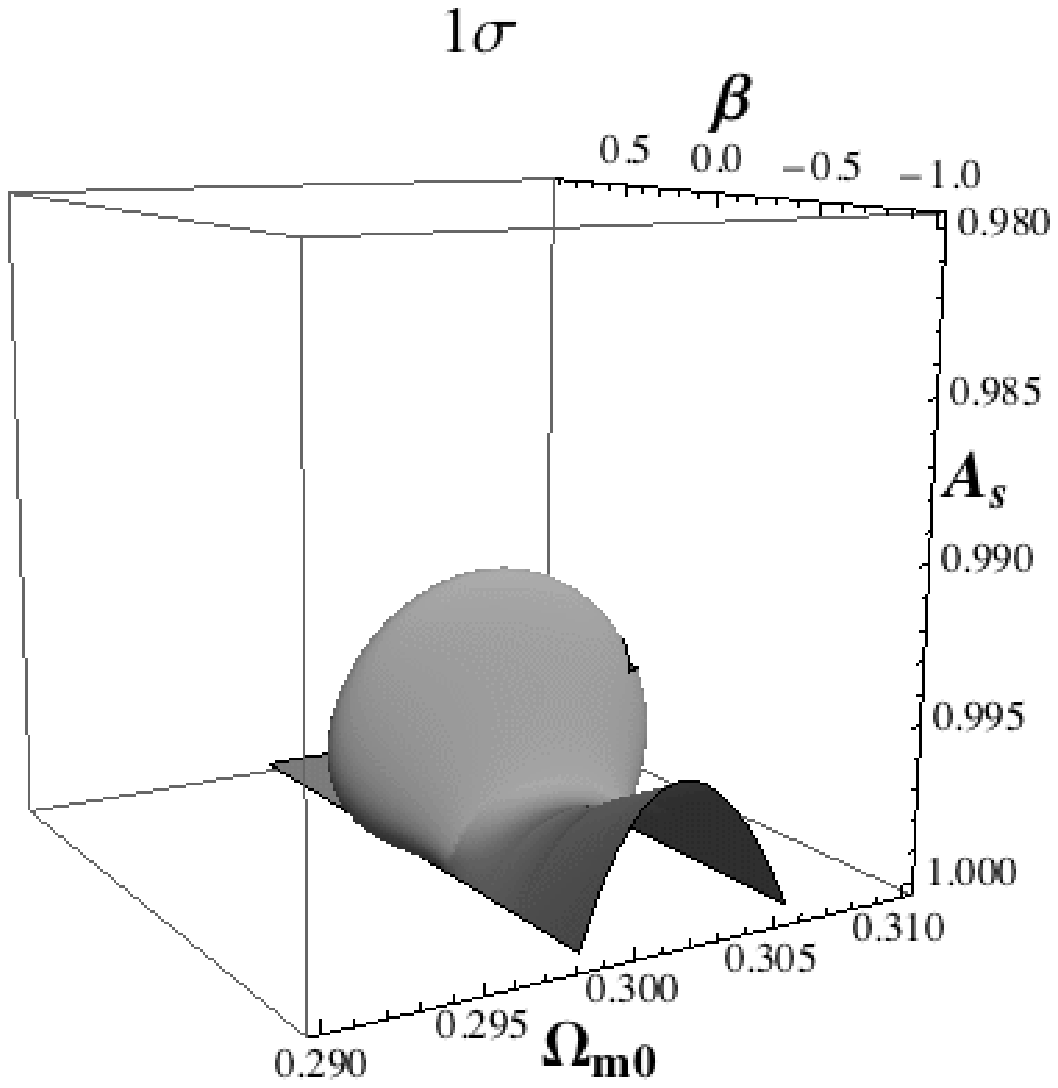}}&
{\includegraphics[width=2.6in,height=2in,angle=0]{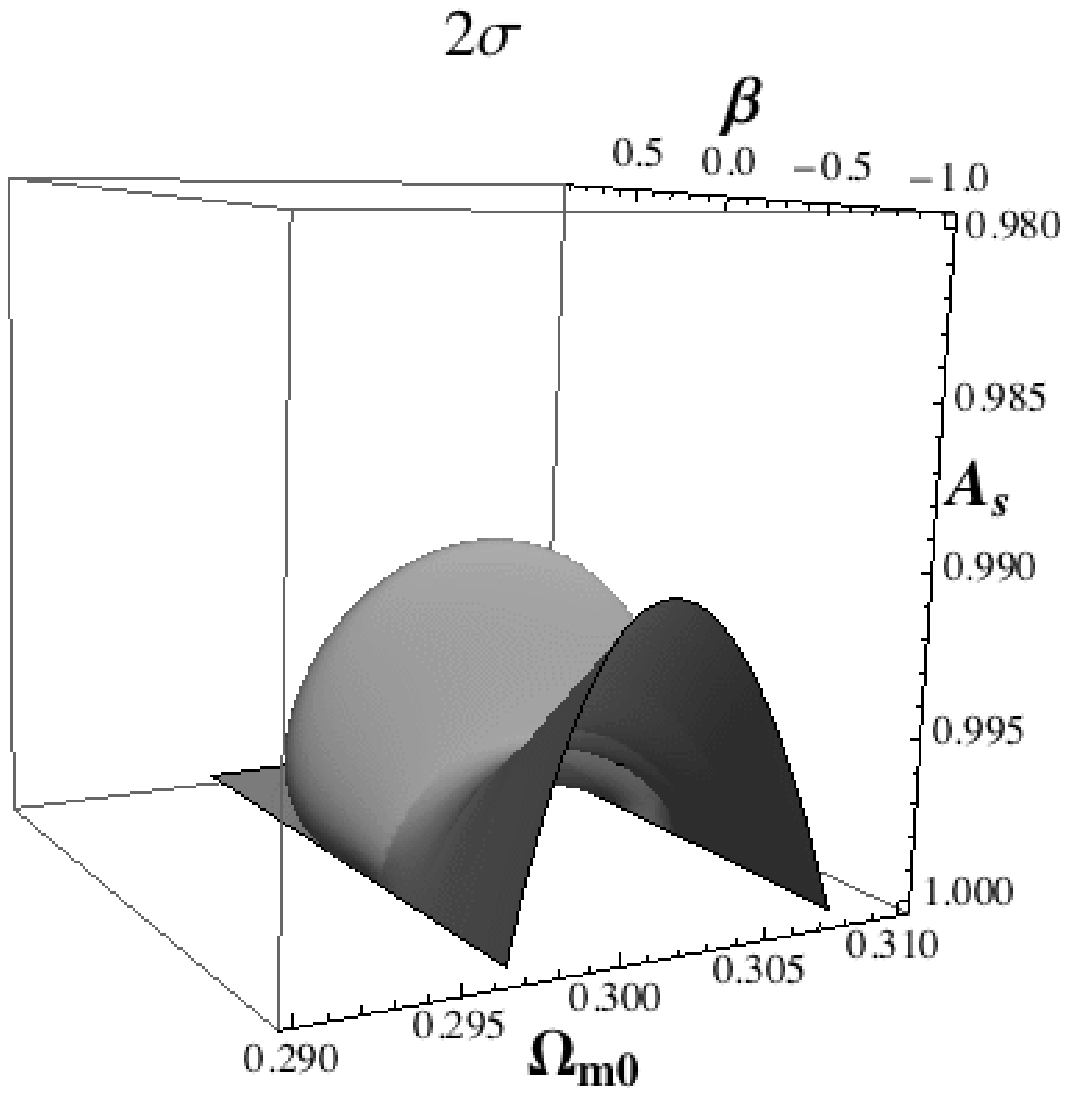}}
\\
\hline
{\includegraphics[width=2.6in,height=2in,angle=0]{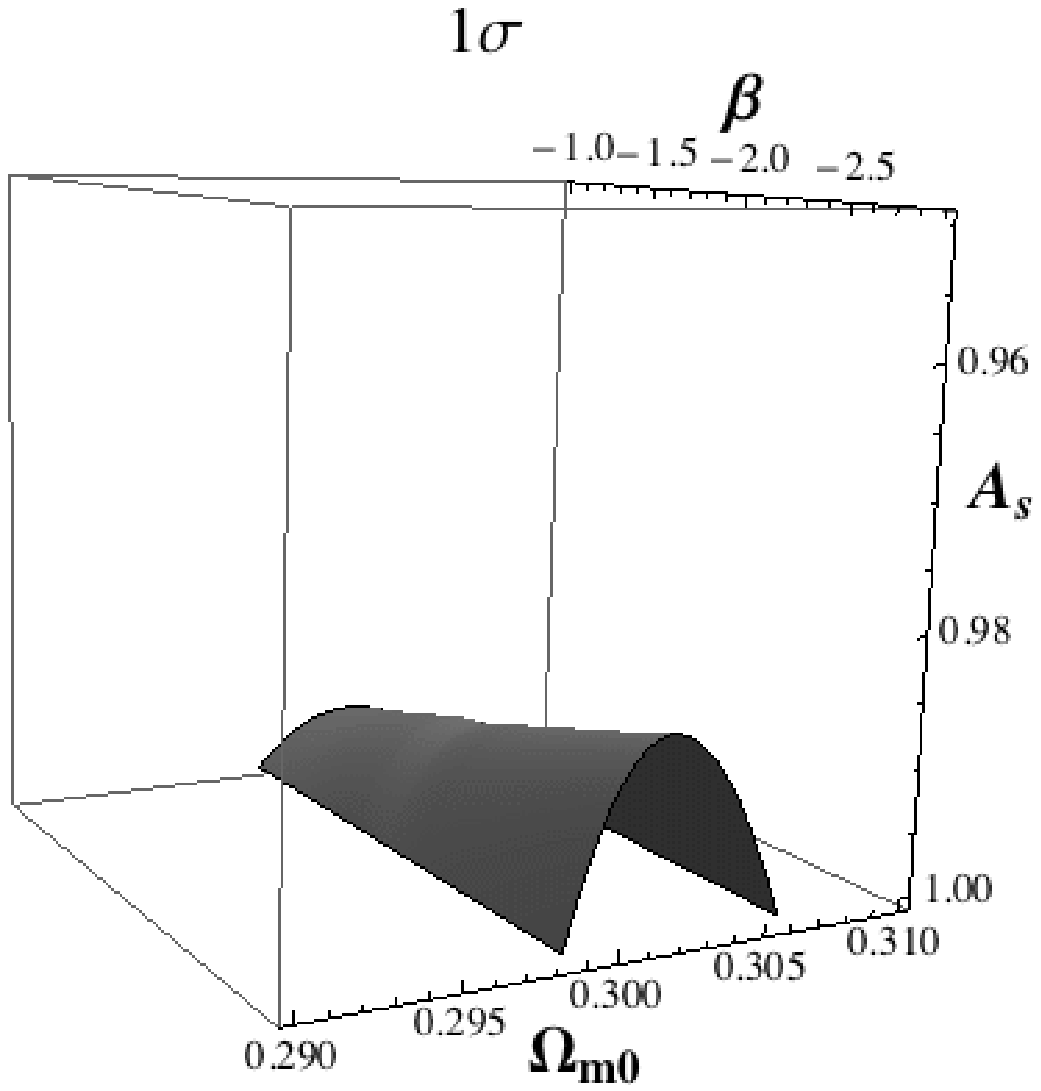}}&
{\includegraphics[width=2.6in,height=2in,angle=0]{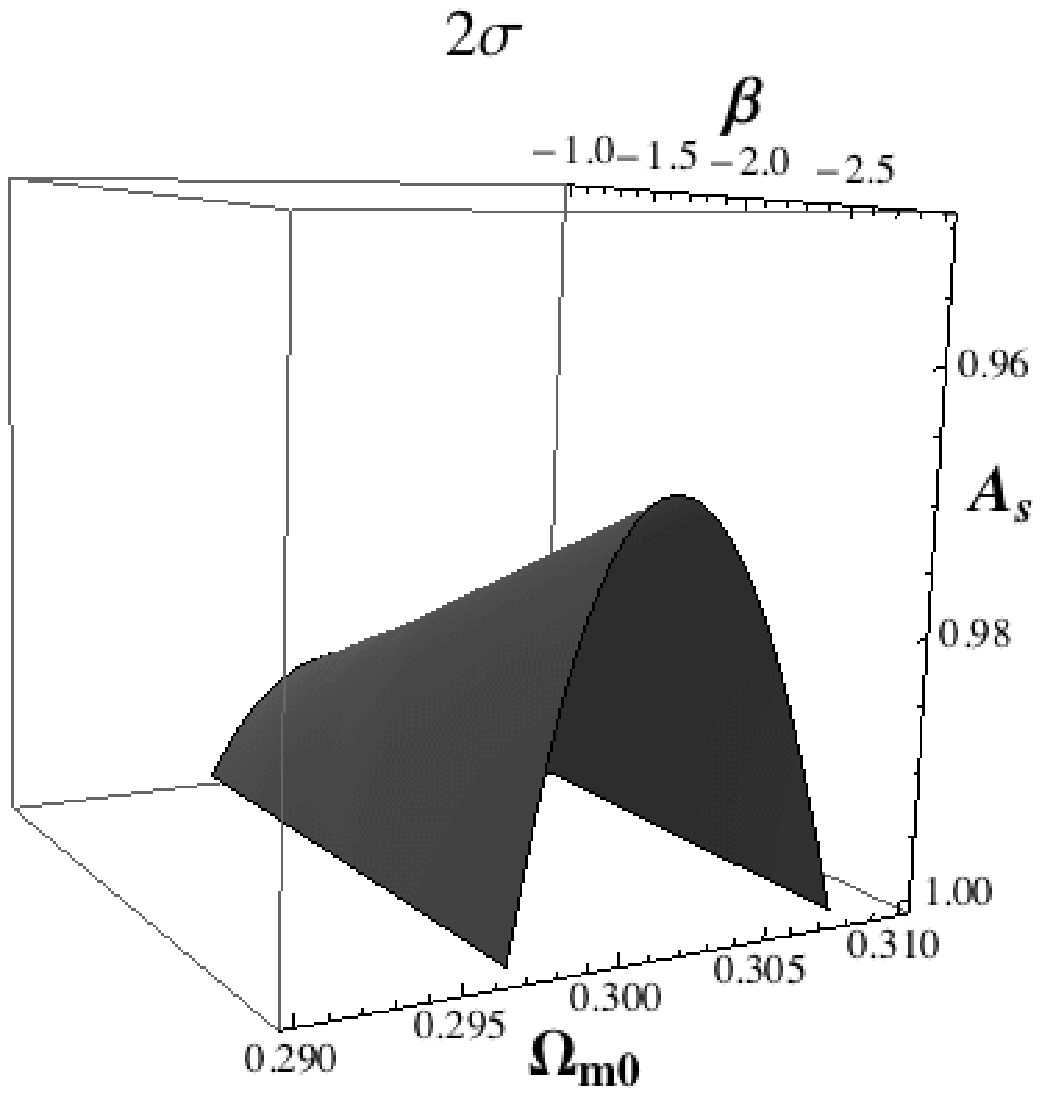}}
\\
\hline
\end{tabular}
\caption{$1\sigma$ and $2\sigma$ confidence contours  in the $\Omega_{m0}-A_{s}-\beta$ space. The contours are plotted with all the observational data mentioned in the text assuming the perturbed dark energy. The upper row is tracker type of models while lower one is for thawer type of models.}
\end{center}
\end{figure*}

Considering the fact that GCG can effectively represent both the thawing as well as the tracking behaviour by suitable parametrization, we use this feature to discriminate between these two models on the basis of Bayesian analysis using the current observational data.
\section{Data}
We consider the Supernovae Type Ia observation which is a direct probe for the cosmic expansion. These observations measure the apparent luminosity of the Supernovae as observed by an observer on earth. It is related to the luminosity distance $d_{L}(z)$ 
defined  as 
\begin{equation}
d_{L}(z) = (1+z)\int_0^z\frac{dz^{\prime}}{H(z^{\prime})}\end{equation}

Using this, we evaluate the distance modulus $\mu$ (which is an observable quantity) as
\begin{equation}
\mu = m-M = 5\log\frac{d_{L}}{Mpc}+25,
\end{equation}
where m and M are, respectively, the apparent and absolute magnitudes of the Supernovae. We consider the latest Union2.1 data compilation (\ Suzuki et al.~(2011)) consisting of 580 data points for the observable $\mu$.

We also use determinations of the cosmic expansion history from red-envelope galaxies by Stern et al. (\ Stern et al.~(2010)). They have obtained a high-quality spectra of red-envelope galaxies in 24 galaxy clusters in the redshift range $0.2< z< 1.0$ using the Keck-LRIS spectrograph. Subsequently they have complemented these with high-quality, archival spectra as obtained by the SPICES and VVDS surveys. With this, we get 12 measurements of the Hubble parameter H(z) at different redshifts. The measurement of the Hubble parameter at present (i.e. $z=0$) is obtained from the HST Key project.

The observational data of SNe Ia and Hubble(H(z)), mentioned above relies on the measurements of cosmological distances at certain redshifts.  On the other hand, the lookback time to high redshift objects, as proposed by (\ Dalal et al. (2001)) depends on the measurement of age of the Universe at a particular redshift.
The test based on the lookback time when
combined with the distance measurements can be very useful to constrain cosmological models.

Lookback time is defined as the difference between the present age of the Universe $(t_0)$ and its age at redshift z, $t(z)$ and is given by (\ Cappzziello et al. (2004))
\begin{eqnarray}
t_L(z,p)=t_0(\theta)-t(z)  \\= \frac{1}{H_0}\left[\int_0^{z} \frac{dz^\prime}{(1+z^\prime)\mathcal{H}(z^\prime,\theta)}\right], 
\end{eqnarray}
where $\theta$ represents the set of model parameters, and  $ \mathcal{H}(z^\prime,\theta)=H(z^\prime,\theta)/H_0$.

The observed lookback time of an object at redshift $z_i$ is defined as 
\begin{eqnarray}
t_L(z_i,\theta)= t_0^{obs}-t(z_i)-t_{inc},
\end{eqnarray}
where $t_0^{obs} $ is the observed age of the Universe. We use the value of $13.6_{-0.3}^{+0.4}$ billion years for $t_0^{obs}$ (\ MacTavish et al (2005)).  The incubation time has been denoted by $t_{inc}$ and it is different for different objects. It is taken to be a nuisance parameter in our study and we marginalize over this parameter (\ Dantes et al. (2007)). The data for lookback time is obtained from (\ Simon et al. (2005)). 

\begin{figure*}
\begin{center}
\begin{tabular}{|c|c|}
\hline
 & \\
{\includegraphics[width=2.6in,height=2in,angle=0]{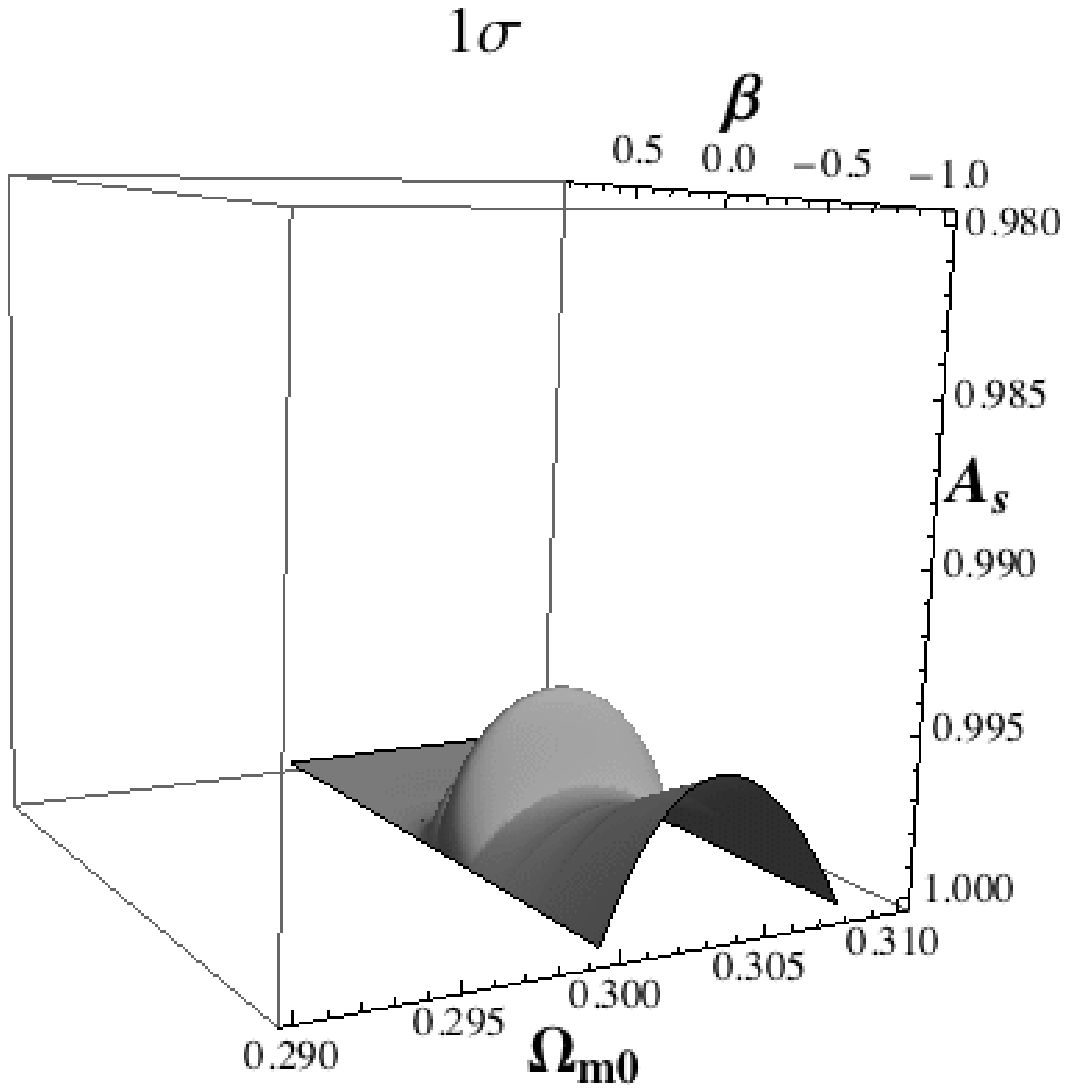}}&
{\includegraphics[width=2.6in,height=2in,angle=0]{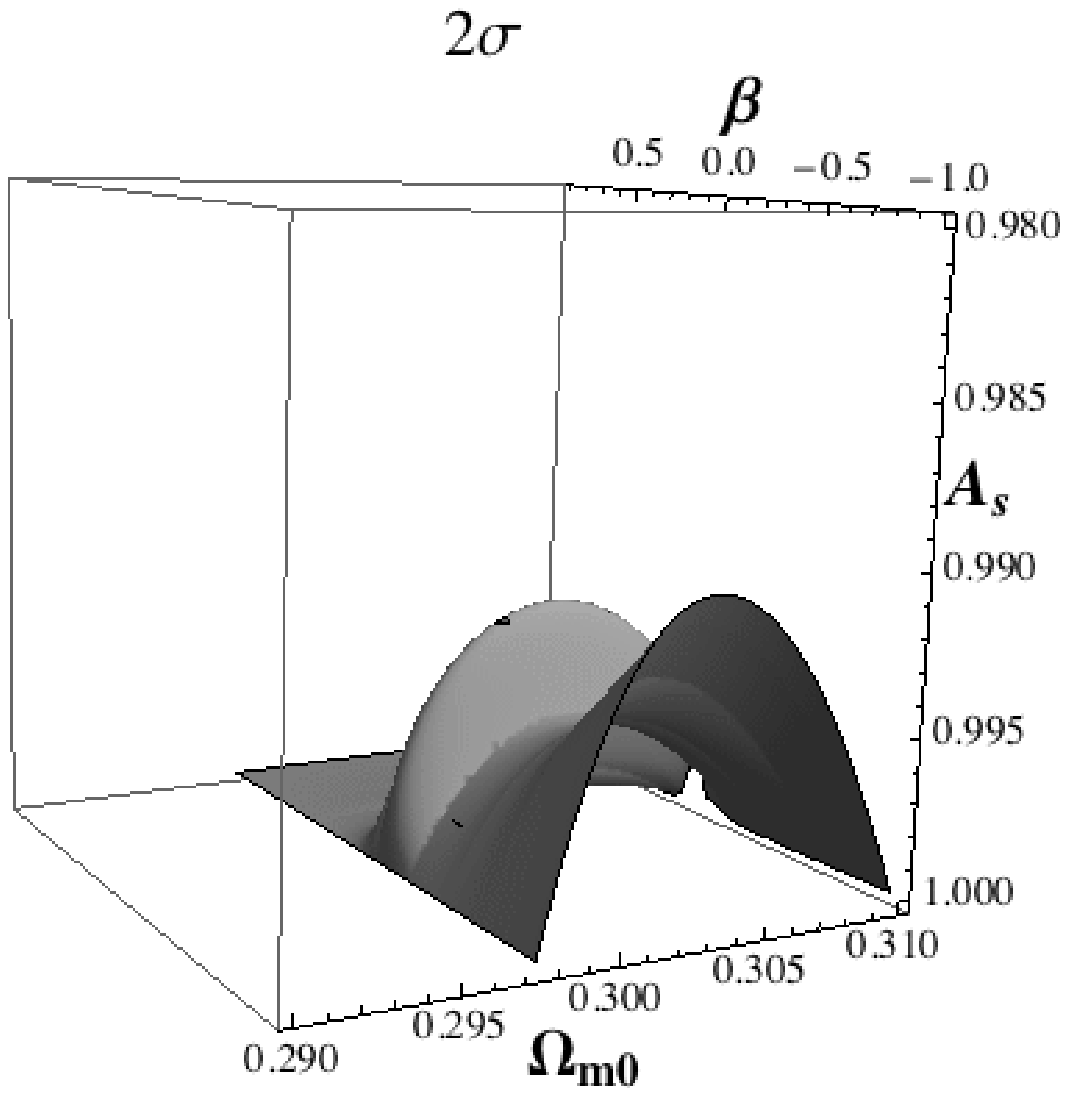}}
\\
\hline
{\includegraphics[width=2.6in,height=2in,angle=0]{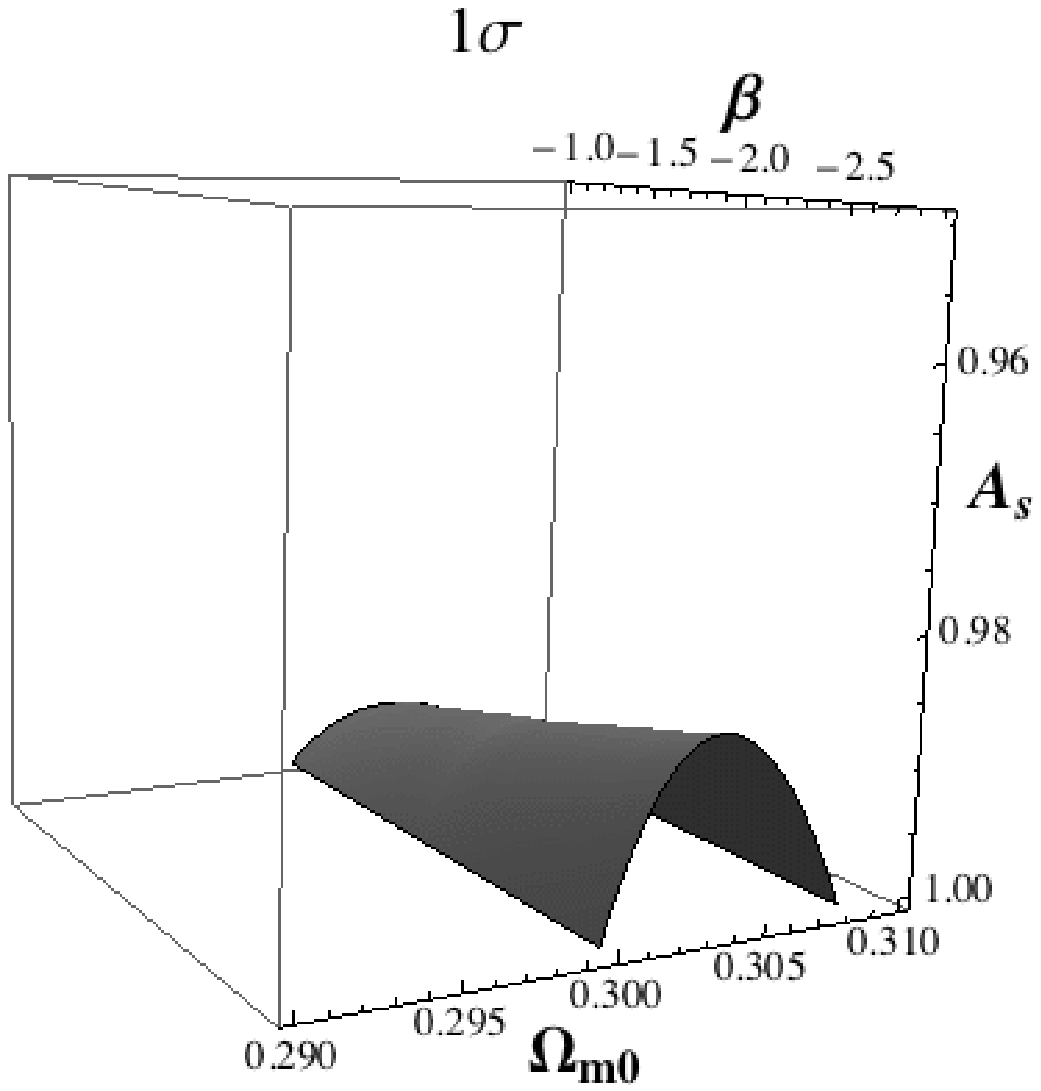}}&
{\includegraphics[width=2.6in,height=2in,angle=0]{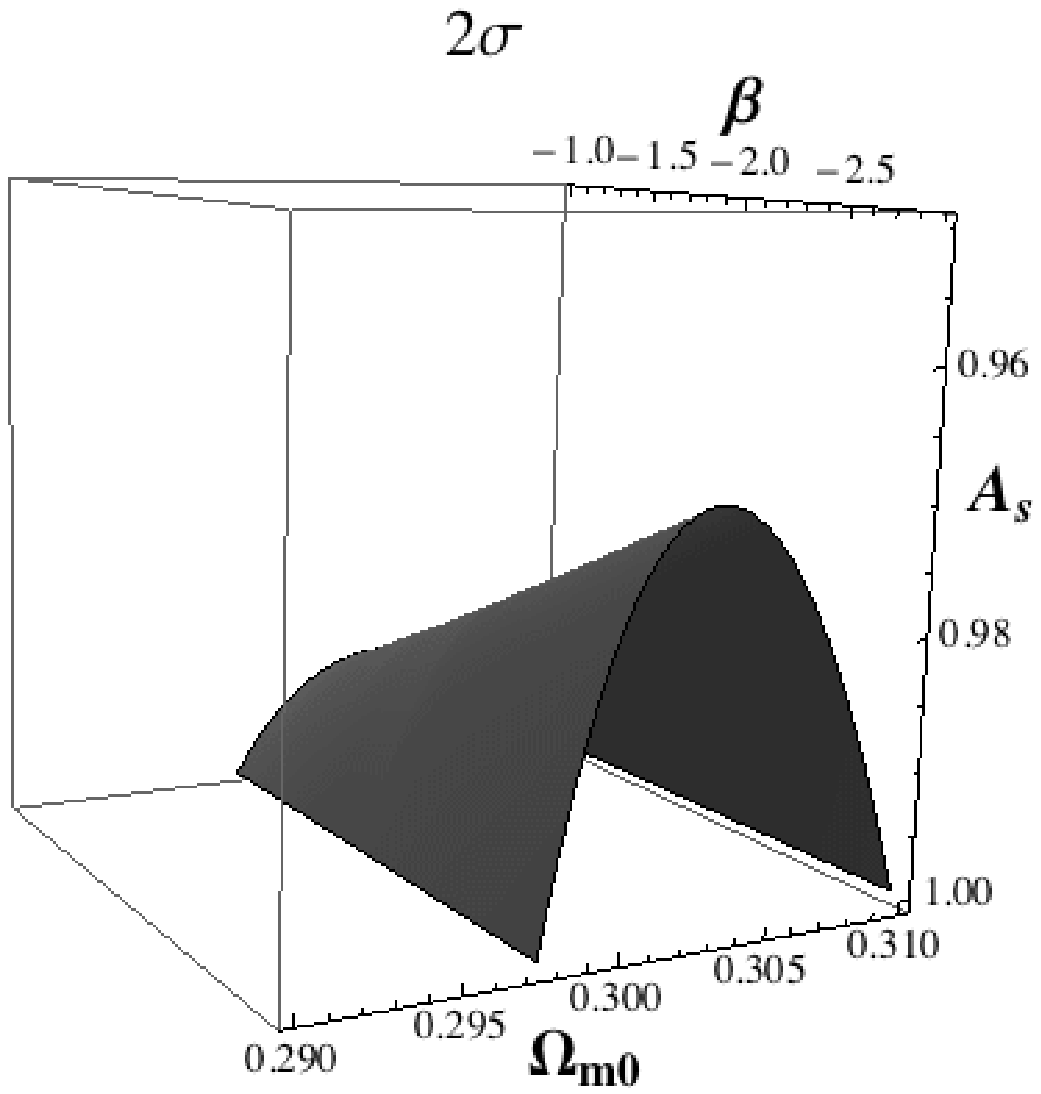}}
\\
\hline
\end{tabular}
\caption{$1\sigma$ and $2\sigma$ confidence contours  in the $\Omega_{m0}-A_{s}-\beta$ space. The contours are plotted with all the observational data mentioned in the text assuming the smooth dark energy. The upper row is tracker type of models while lower one is for thawer type of models.}
\end{center}
\end{figure*}

We further use the measurement of CMB anisotropy  by WMAP-7 observations (\ Komatsu et al.~(2011)). Using the publicly available code  
CAMB (\ Lewis et al.~(2000)), we calculate the expected CMB anisotrpy spectrum for GCG as a function of its two model parameters $A_{s}$ and $\beta$  and the cosmological parameter $\Omega_{m0}$.  To incorporate the dark energy perturbation, we use publicly available PPF Module for CAMB (\ Fang et al~(2008a), \ Fang et al.~(2008b)). In this case, GCG fluid is used as a dark energy with equation of state given in eqn (4) and the speed of sound $c_{s}^2 = 1$ as this fluid represents a parametrization of the scalar field models ( For detail calculation and relevant equations in PPF module, see (\ Fang et al.~(2008b))). 

 The other cosmological parameters are equated to their best fit values as derived from the data from WMAP-7.  We compare our theoretical results with the data and do the likelihood analysis using the likelihood code provided by the WMAP team (\  Komatsu et al.~(2011) ). 

The growth of large scale structures is another useful observational probe to distinguish between different dark energy models.  The growth rate of large scale structures is calculated from matter 
density perturbation $\delta=\delta\rho_m/\rho_m$ in the linear regime and subhorizon scales $(\frac{k^2}{a^2}>>H^2)$.  On these scales where one measures the growth, one can safely ignore the dark energy perturbations as it only affects the fluctuations comparable to Hubble scales. The effect of dark energy is only through the background expansion and one can use the Newtonian limit for the evolution of the matter density fluctuation satisfying the equation:

\begin{equation}
\ddot{\delta}+2\frac{\dot{a}}{a}\dot{\delta}-4\pi G\rho_{m}\delta=0 .\\
\end{equation}

The growth factor $f$ is related to the linear density contrast $\delta$ by $f \equiv \frac{dln\delta}{dlna}$. The compilation of data for growth factor measurements from different redshift surveys can be found in (\ Gupta et al.~(2011)) ( See also references therein).
\section{Results}
Using observational data mentioned above we calculate the joint likelihood for the GCG model as:
\begin{equation}
-2\log {\cal L} = - 2\log ({\cal L}_{sn} \times {\cal L}_{H} \times {\cal L}_{look} \times {\cal L}_{wmap} \times {\cal L}_{gr}).
\end{equation}

Before calculating the Bayesian Evidence for thawing type and
  tracker type of models, we maximise the likelihood function and
  calculate the allowed region in the parameter space of $A_s, \beta$
  and $\Omega_{m0}$ for both thawing type and tracking type of models. The results are shown in Figure (2) and Figure (3). In these figures we show the 3-D contours in the $A_{s}-\beta-\Omega_{m0}$ region. For the tracker type of models ( shown in Figure (2 and 3)), it is clear that upto $2-\sigma$ confidence level, the allowed behaviours are extremely close to C.C. This is apparent from the bound on $A_{s}$ which varies negligibly from $A_{s}=1$, $
\Lambda$ behaviour. The constraint on $\Omega_{m0}$ parameter is also extremely tight, $0.295\leq\Omega_{m0}\leq 0.31$. Also, in this case, all the values of $\beta$ are allowed. But for two specific values of $\beta$, the allowed value of $A_{s}$ is maximum. These are $\beta=-1$ for which GCG behaves as a DE with constant equation of state and for $\beta = 0$ for which GCG itself behaves as a combination of $\Lambda$ and matter fluid. For $\beta >0$ only $A_{s} = 1$ is allowed which is similar to C.C.

For thawing case ( shown in Figure (2 and 3)), the variation of $A_{s}$ parameter is marginally higher than the tracking case, but here also the constraint on $\Omega_{m0}$  is extremely tight. Also in this case, all values of $\beta$ are allowed, but as one increases $\beta$, the allowed values for $A_{s}$ get decreased. 

To summarise, with all the data, the GCG parametrization puts  extremely tight constraint on the dark energy behaviour and the cosmological parameter $\Omega_{m0}$.

We now proceed to substantiate this conclusion. In order to do this we calculate the Bayesian Evidence for both the thawing as well as tracking behaviour. This is defined as (\ Liddle et al.~(2006))
\begin{equation}
E = \int {\cal L}(\theta) P(\theta) d\theta,
\end{equation} 
where $\theta$ represents the set of model parameters, ${\cal L}$
represents the Likelihood function and  $P(\theta)$ is the prior
probability distribution for parameter $\theta$.  In our case, we have
three parameters $A_{s}, \beta$ and $\Omega_{m0}$. As we have no prior
knowledge about these parameters, we assume uniform prior for these
three parameters.We get a thawing behaviour for $1+\beta < 0$ and tracking for $1+\beta > 0$. For the thawing case we assume that $\beta$ is uniformly distributed in the range $-3<\beta\leq-1$ while for tracking case we assume that $\beta$ is uniformly distributed in the range $-1\leq\beta< 1$. $A_{s}$ fixes the equation of state for dark energy at present ($A_{s} = - w_{de}(z=0)$) and we assume that it is uniformly distributed between $0.8$ and $1$ for both thawing as well as tracking case.  We also assume that for both the cases, $\Omega_{m0}$ is uniformly distributed between $0.2$ and $0.4$.

According to Jeffrey's interpretation (\ Jeffreys(1998)), if $\Delta \ln E$ between $1$ and $2.5$, it is a significant evidence for a model with higher $E$ while the same between $2.5$ and $5$ is a strong to very strong evidence. If $\Delta ln E$ is more than $5$, the model with higher $E$ is decisively favoured.  

We calculate $\Delta ln E$  assuming thawing ($1+\beta < 0$) and tracking ($1+\beta > 0$) as two different models. We get the following results:{ \bf
\begin{equation}
\ln E_{thaw} - \ln E_{trac} = 0.34\hspace{2mm} with ( sn+hub+look).
\end{equation}

\begin{equation}
\ln E_{thaw} - \ln E_{trac} = -0.55  \hspace{1mm}with(sn+hub+look+gr+wmap)
\end{equation}
\noindent
\hspace{30mm} with D.E. perturbation

\begin{equation}
\ln E_{thaw} - \ln E_{trac} = 1.62  \hspace{1mm}with(sn+hub+look+gr+wmap)
\end{equation}

\noindent
\hspace{30mm} without D.E. perturbation }

\vspace{5mm}
\noindent
From these results, it is clear that although observational data from
the smooth background cosmology cannot discriminate between thawing
and tracking behaviour, when these are complemented with the
information from the inhomogenous universe but assuming uniform dark
energy distribution there is significant evidence in favour of thawing behaviour ( with odds ratio 5:1). However, when we take into account the perturbation in the dark energy component while calculating the CMBR spectra, this preference over thawing type of models is lost and there is no Bayesian evidence to prefere any type of behaviours.

 We should also mention that although we choose a specific prior for $A_{s}$ and $\Omega_{m0}$ as mentioned above, the results are only weakly sensitive to these priors. In order to verify this we have varied these priors and we found that the above results remain almost the same. This makes our result sufficiently robust.

\section{Discussion and Conclusions}
 Several observed features of the Universe depend on the precise
  nature of cosmological expansion. Broadly the evolution of Universe with a dark energy can be classified in two categories. In one case the DE equation of state is frozen at the $w=-1$ in the early universe due to large Hubble friction but in the late time as the Hubble friction eases out, the equation of state slowly thaws out of this frozen state. This is so called thawing behaviour and can be modelled by the scalar field with slow-roll potentials. In the other case, the DE equation equation of state mimics the background  matter in the early universe and in the late time it starts behaving like a C.C. This type of behaviour is called tracker type and can be mimicked by the scalar fields with steeper potentials. Here we parametrize these two overall behaviours by the GCG type of equation of state and investigate whether the observational data prefer any one of them.  

Towards this end we have first used data pertaining to the background cosmology. In particular, we have considered three types of data:  (i) Supernova Type 1A, (ii) Hubble parameter evolution with redshift, and (iii) Lookback time of Galaxies. 
For the Supernova Data we have used the Union2.1 data. The data from
Keck-LRIS, SPICE and VVDS has been used for the evolution of hubble
parameter while for the present redshift hubble parameter HST Key
project has been used. For the look back time calculation, the incubation time of the galaxies has been marginalized over as we do not have definitive information about it.  

We have then complemented these with the information from inhomogeneous universe. For this we have used the results of CMB Anisotropy (as given in the WMAP-7 data) and the growth of large-scale structures.  

Our results indicate that if we take into account only the smooth
background cosmology, then the data is insufficient to give a clear
verdict either way. On the other hand, if we complement this with the
information from inhomogeneities in the Universe then there seems to
be a preference for the thawing model if one does not consider the
dark energy perturbations while calculating the CMBR spectra. In terms of Bayesian evidence, this preference for thawing behaviour is significant. However, we find that incorporating the dark energy perturbation to calculate the CMBR spectra, this preferance for thawing type of behaviour is lost.

We stress that the GCG cannot represent the full scalar field dynamics for tracker type of models in particular and this weakens our conclusion. Nevertheless, our analysis is the first one to investigate whether observational data prefer a thawing or tracking type of behaviour.

\section{Acknowledgement}
A.A.S. and T.R.S. acknoweldge the financial support provided by C.S.I.R. Govt. of India through the research grant (Grant No:03(1187)/11/EMR-II). S.T is fully funded by C.S.I.R. Govt. of India. S.T. and T.R.S. acknowledge facilities provided by Inter University for Astronomy and Astrophysics, Pune, India through IUCAA Resource Centre (IRC) at Department of Physics and Astrophysics, University of Delhi, New Delhi, India. S.T. acknowledges the computational facilities provided by the CTP, JMI, New Delhi, India.  Part of the numerical computations were performed using the Cluster computing facility at the Harish-Chandra Research Institute,Allahabad, India (http://cluster.hri.res.in/index.html).  

We thank the LAMBDA project  for providing the data and likelihood code for WMAP7 (http://lambda.gsfc.nasa.gov/product/map/current/). We thank A. Lewis and A. Challinor for the publicly available CAMB code. We also thank W. Fang for the publicly available PPF module for CAMB.

\end{document}